\documentclass[10pt,twocolumn,letterpaper]{article}

\usepackage{cvpr}
\usepackage{times}
\usepackage{epsfig}
\usepackage{graphicx}
\usepackage{amsmath}
\usepackage{amssymb}
\usepackage[percent]{overpic}
\usepackage{boldline}
\usepackage{multirow}
\usepackage{float}
\usepackage{bm}
\usepackage{latexsym}
\usepackage{amsthm}
\usepackage[ruled]{algorithm2e}
\usepackage{subfigure}
\usepackage{enumerate}
\usepackage{enumitem}
\usepackage{bbding}
\usepackage[table,xcdraw]{xcolor}

\usepackage[breaklinks=true,bookmarks=false]{hyperref}

\cvprfinalcopy 

\def\cvprPaperID{984} 

\def\R{\mathbb R}

\def\cS{\mathcal S}
\def\cU{\mathcal U}

\def\x{\mathrm x}
\def\a{\mathrm a}
\def\t{\mathrm t}
\def\v{\mathrm v}
\def\G{\mathrm G}
\def\P{\mathrm P}
\def\E{\mathrm E}
\def\GRU{\mathrm{GRU}}
\def\ATTN{\mathrm{ATTN}}
\def\Th{\bm{\Theta}}

\renewcommand{\u}{\bm{\mathrm{u}}}
\newcommand{\bb}{\bm{\mathrm{b}}}
\newcommand{\h}{\bm{\mathrm{h}}}

\newtheorem{defPartial}{Definition}

\SetKwInput{KwInit}{Initialize}
\ifcvprfinal\fi
\begin{document}
\setlength\abovedisplayskip{3pt}
\setlength\belowdisplayskip{3pt}


\title{M2R2: Missing-Modality Robust emotion Recognition framework with iterative data augmentation}

\author{
Wang Ning \\
Xi'an Jiaotong University \\
{\tt\small @}
\and
a\\
a\\
{\tt\small @}
}

\maketitle

\begin{abstract}
This paper deals with the utterance-level modalities missing problem with uncertain patterns on emotion recognition in conversation (ERC) task. 
Present models generally predict the speaker's emotions by its current utterance and context, which is degraded by modality missing considerably.
Our work proposes a framework Missing-Modality Robust emotion Recognition (M2R2), which trains emotion recognition model with iterative data augmentation by learned common representation. 
Firstly, a network called Party Attentive Network (PANet) is designed to classify emotions, which tracks all the speakers' states and context. Attention mechanism between speaker with other participants and dialogue topic is used to decentralize dependence on multi-time and multi-party utterances instead of the possible incomplete one.
Moreover, the Common Representation Learning (CRL) problem is defined for modality-missing problem. Data imputation methods improved by the adversarial strategy are used here to construct extra features to augment data. 
Extensive experiments and case studies validate the effectiveness of our methods over baselines for modality-missing emotion recognition on two different datasets.
Impact Statement
\end{abstract}

\section{Introduction}

\begin{figure*}[htbp]
\begin{center}
\includegraphics[width=0.7\linewidth]{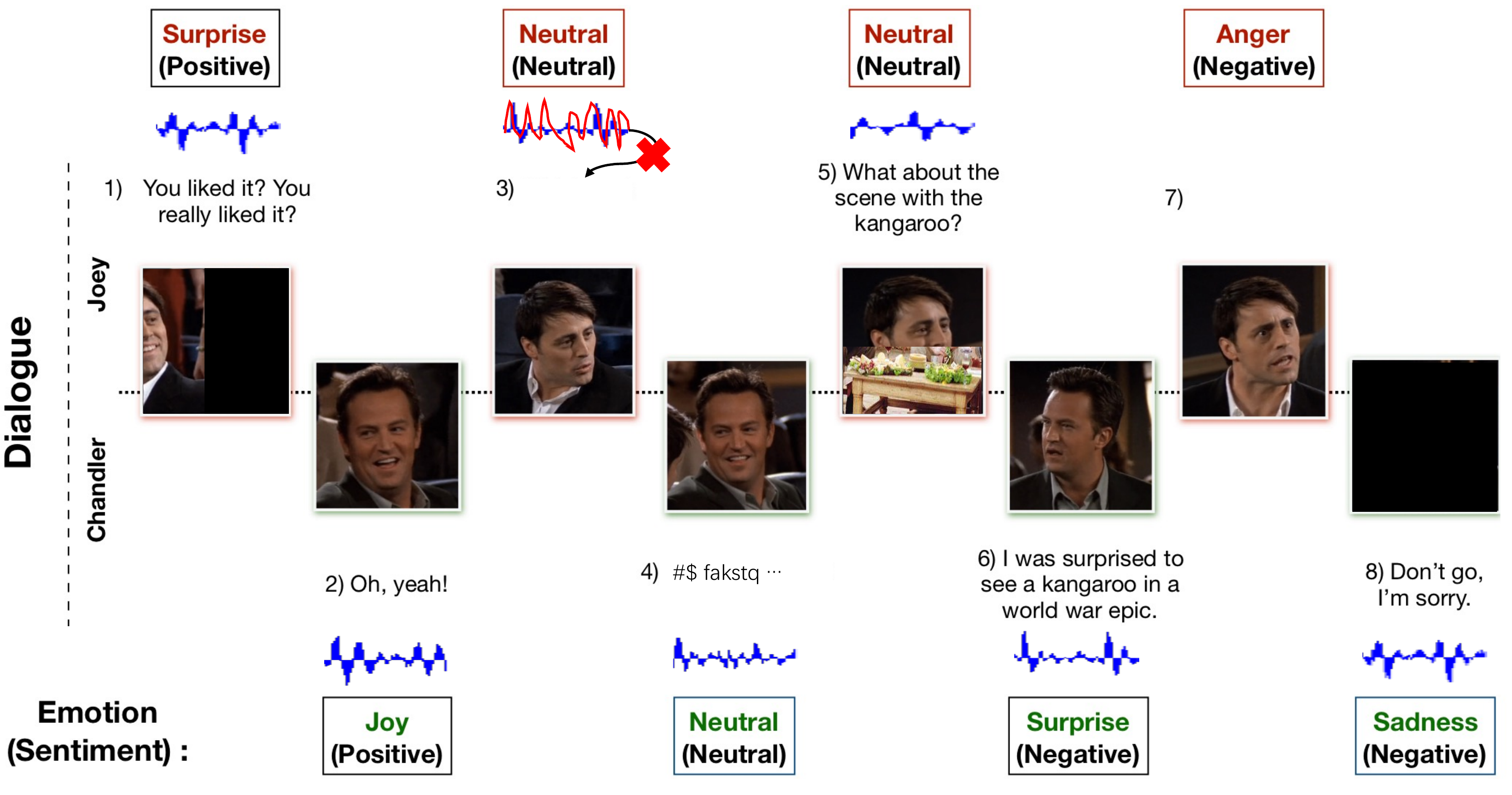}
\end{center}
   \caption{Possible causes of data missing at utterances in a conversation: 1) target face out of image boundary; 3) audio corrupted by noise; 4) fail to transform audio to text; 7) audio sensor fails; 8) camera fails.}
\label{fig:Intro}
\end{figure*}

Emotion recognition in conversation (ERC) has now attracted increasing attention as its potential ability to mine and analyze the connections from the heavily incremental volume of publicly available conversational data\cite{reviewercsentimentai}\cite{reviewhci}. Furthermore, ERC can be used as a tool for psychological analysis in healthcare systems\cite{defaffective} and human-interacted emotion-aware conversation that requires an understanding of the user's emotions \cite{reviewerc}\cite{reviewreallife}.

Traditional machine learning methods according to textual features such as Naïve Bayes classifier\cite{bayes}, NLTK vader\cite{vader}, etc, have been proposed for ERC task \cite{reviewml}. As deep learning develops, many neural networks models are employed to classify the emotion at each utterance. \cite{autoencoder} uses autoencoder to detect multiple emotions in conversation transcripts. Graph convolutional networks are exploited by \cite{igcn} and \cite{gcn} to model the inter-relationship between conversation topic and its different parties for better emotion representations. \cite{hierarchical} employs LSTM cells and Bert hierarchically for utterance context considering the sequential essence of the conversation, which was followed by \cite{bclstm} firstly using recurrent neural network (RNN) to recognize emotions. Moreover, many NLP techniques and ideas are introduced in ERC. \cite{cosmic} incorporates different elements of commonsense and builds upon them to learn interactions between all conversation parties, and \cite{knowledge} proposes context-aware affective graph attention mechanism to interpret contextual utterances hierarchically.

While solving the ERC task by unimodal (mostly textual) data is not enough because not only that the emotions are subjective mixed feelings \cite{reviewreallife}, but also the labeling difficulty in real-life conversation scenarios. Through multiple channels, humans express emotions and feelings (such as voice speech, textual meaning, facial expression, body gestures, movements) provides multimodal data. Multimodal emotion recognition systems are studied alongside unimodal solutions as they offer higher accuracy of classification \cite{reviewmultimodal}. Many systems consider multimodal data for ERC. \cite{cmn}, \cite{icon} and \cite{dialoguernn} introduce attention mechanism in conversation context extraction by pooling all proceeding global states related to concatenated utterance modalities. To better fuse multi modalities under different representation spaces, \cite{interactive}, \cite{fuzzy}, \cite{crosstransformer} propose kinds of novel cross-modality modules to capture intra- and inter-modal influence.

In the real world, however, data often suffer from utterance-level incompleteness \cite{overviewpartial} with various patterns, which greatly degrades the integration and complement between different modalities. For the ERC task, there are various causes for desired data to be temporarily missing: sensors can fail so that the corresponding signal is no longer available, or its outcome may be corrupted by noise. In another situation, faces with obvious expressions may be blocked or move drastically and then disappear from the view of the sensor \cite{overviewmissing}. 

Few works focus on the modality-missing problem at utterance level. Typical data imputation algorithm \cite{missing1}\cite{missing2}\cite{missing3} are aimed at separate, non-sequential data. \cite{TrainingST} pad missing modalities by zeros or repeated data with advanced training strategies. \cite{MMIN} learns robust joint multimodal representations by bidirectional encoders with cascade-connected residual imagination module. \cite{SMVAE} performs infinite imputations by integrating all modalities, which was followed by \cite{MMER} which leverages several classifiers under all combinations of different modalities. By building features reconstruction module, \cite{SMIL} employs Bayesian neural network to employ data uncertainty as feature regularization.

This paper deals with the utterance-level modalities missing problem with uncertain patterns on emotion recognition in conversation (ERC) task. 
We propose a novel framework Missing-Modality Robust emotion Recognition (M2R2), which trains emotion recognition model with iterative data augmentation by learned common representation. 

We first introduce Party Attentive Network (PANet), which tracks all the speakers' states and context. Attention mechanism between speaker with other participants and dialogue topic is used to decentralize dependence on multi-time and multi-party utterances instead of the possible incomplete one.
Moreover, we define the Common Representation Learning (CRL) problem for modality-missing data augmentation. Data imputation methods improved by the adversarial strategy are used here to construct extra features. Extensive experiments and case studies validate the effectiveness of our methods over baselines for modality-missing emotion recognition on two different datasets.
The contributions are summarized as follows:
\begin{enumerate}[itemsep=5pt, topsep=5pt, parsep=0pt, itemsep=5pt]
    \item We propose a novel M2R2 framework with iterative data augmentation on missing-modality datasets, which alleviate the performance decline caused by modalities incompleteness greatly. 
    \item To better alleviate the degradation on ERC task caused by modality missing at utterance level, this paper designs PANet to utilize attention mechanism in speaker with other participants and dialogue topic.
    \item A new CRL task is defined to learn extra features for data augmentation. Adversarial strategy, WGAN, is employed to generate extra features better.
\end{enumerate}

\section{Problem Statements}\label{sec:Pdefinition}
Suppose a multimodal conversation scene with $T$ length, let
$\bm{\x}^{\a}=[x^{\a}_1,\dots,x^{\a}_T], \bm{\x}^{\t}=[x^{\t}_1,\dots,x^{\t}_T], \bm{\x}^{\v}=[x^{\v}_1,\dots,x^{\v}_T]$ define audio, text and visual representations respectively. And there are $Q$ parties/speakers $q_1, q_2, \dots, q_Q $ in this conversation. ERC task is to predict the emotion label $e_t^{q_{I\left(t\right)}}$ at time $t$, where $I(t) \in {1,2,\dots,Q}$ indicates the mapping from time $t$ to index of current party. At each time $t$, the constituent utterance $\u_t=\left(x^{\a}_t \oplus x^{\t}_t\oplus x^{\v}_t\right) \in \R^{D_{\a}+D_{\t}+D_{\v}}$ is uttered by $q_{I(t)}$, where $\oplus$ means concatenation here, and this multimodal representations $\u_t$ is extracted as described in section \ref{sec:feature}.
Current systems use GRU cells \cite{gru} to update hidden state and result representation. At time step $t$, the GRU cell computes a new hidden state $h_{t}$ by $\GRU(x_t, h_{t-1})$ as follows:
\begin{equation}
    \begin{split}
    z_t &= \sigma(W_z x_t+ U_z h_{t-1}), \\
    r_t &= \sigma(W_r x_t+ U_r h_{t-1}), \\
    \Tilde{h}_t &= \mathrm{tanh}(W x_t + U(r_t\circ h_{t-1})), \\
    h_t &= (1-z_t) \circ h_{t-1} + z_t \circ \Tilde{h}_t,
    \end{split}
\end{equation}
where $h_{t-1}$ is previous hidden state and $x_t$ is current input.$W_z, U_z, W_r,U_r,W,U$ server as trainable parameters which make $\GRU$ efficient.

Given the lack of several modality at different time, a new data piece $\{ \cS_t, e_t \}_{t=1}^T$ can be obtain as the subset of the full utterance $\cU_t=\{ \u_t^{m}\}_{m=1}^M$ (i.e, $\cS \subseteq \cU$). Here $M$ means the number of modalities and equals to $3$ in ERC task. If a modality $m_0$ lacks during whole conversation, the new data piece will be $\{\cU_t-\u_t^{m0},e_t\}_{t=1}^T, \forall t \in \{1,2,\dots,T\}$. Consider that a conversation cannot continue if all modalities lacks at the same time, there must be available one for all $t$. Accordingly, we define missing rate as $\eta={\sum_t\sum_m{S_t}}/{\sum_t\sum_m{\u_t^m}}$. Fig.\ref{fig:Intro} describes the possibilities.

For the case of data incompleteness, we define Common Representation Learning task as follows:
\begin{defPartial}
\textbf{Common Representation Learning (CRL)} \quad Given the training set $\{\cS_t, y_t \}_{t=1}^T$ that can be obtain as the subset of the full utterance $\cU_t=\{ \u_t^{m}\}_{m=1}^M$ (i.e, $\cS \subseteq \cU$), CRL aims to learn common representation $\{\h_t\}_{t=1}^T$ of $\{\cS_t\}_{t=1}^T$ under arbitrary missing pattern.
\end{defPartial}
The best common representation can be obtained by maximizing $p(y,\cS|\h)$ as
\begin{equation}
    \h^* = \arg_{h\in\R^{D_\mathcal{H}}}\max{p(y,\cS|\h)},
\end{equation}
where $D_\mathcal{H}$ is the dimension of common representation. To employ log likelihood, we assume that
\begin{enumerate}[itemsep=5pt, topsep=5pt, parsep=0pt, itemsep=5pt]
    \item [(1)] Each modality is conditionally independent;
    \item [(2)] Utterance are independent and identically distributed (i.i.d).
\end{enumerate}

This assumption gives a probability aspect for CRL. Under the above assumption $(1)$, the probability of label $y$ can be viewed as 
\begin{equation}
    p(y,\cS|\h) = p(y|\h)p(\cS|\h),
\end{equation}
where $p(\cS|\h)=p(\u^1|\h)p(\u^2|\h)\dots p(\u^M|\h)$ and $y$ is treated as another semantic modality, with $M$ being the number of modalities. Here we omit the time token for simplification. 

\section{Methodology}
\begin{figure*}[htbp]
\begin{center}
\includegraphics[width=\linewidth]{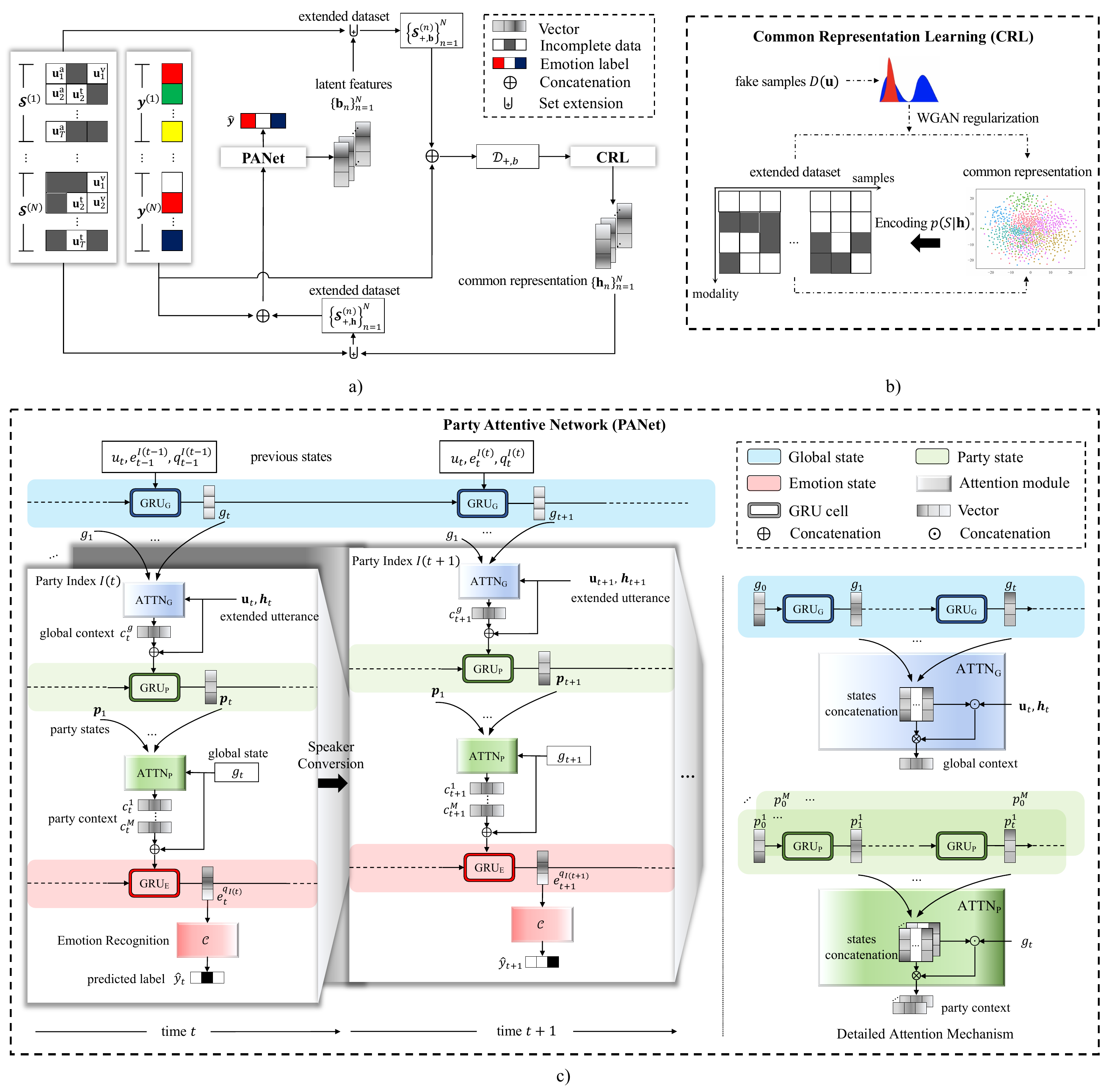}
\end{center}
    \vspace{-1em}
    \caption{Illustration of Missing-Modality Robust emotion Recognition framework on ERC task. a) The procedure of M2R2. PANet and CRL are learnt iteratively on extended dataset with each other's latent features, where PANet model predicts emotion label. b) CRL procedure. Given the modality-missing data, the encoder maps common representation to align modalities. WGAN regularization is used here to improve construction. c) PANet structure. Global GRUs and party, emotional GRUs for all parties are employed to tract and update states. Emotion recognition module exploits multi-layer perception to predict emotion labels. Detailed attention mechanism module is shown in right column.}
    \vspace{-1em}
\label{fig:ep}
\end{figure*}

In this work, we propose a Missing-Modality Robust emotion Recognition framework (M2R2) for the ERC task. Fig.\ref{fig:ep} illustrates the overview and detailed module structures. M2R2 updates PANet and CRL with iterative data augmentation by each other's latent representations.

\subsection{M2R2 Framework}

At training stage, PANet model and CRL are updated by turns with other's output until the accuracy converges. We first initialize network parameters $\Th_e,\{ {\Th_{\u}^m} \}_{m=1}^M$ and common representation $\{ \bb_n \}_{n=1}^N$ with random values. And initializing $ \{ \h_n \}_{n=1}^N $ as zero vectors to extend data set $\mathcal{D}$ to $\mathcal{D}_{+,h}$ primitively for PANet model. 
\vspace{-1em}
\begin{algorithm}[hbp]\label{algo:ep}
    \caption{M2R2 Framework}
    \label{algorithm:test}
    \textbf{Training Stage}\par
    \KwIn {Incomplete training conversation dataset: $\mathcal{D}=\{ (\bm{\mathcal{S}}^{(n)}, \bm{y}^{(n)}) \}_{n=1}^N$ \quad \quad Hyper-parameters: $\lambda_{\mathrm{R}},\lambda_{\mathrm{P}},\lambda_{\mathrm{A}},\lambda_{\mathrm{G}},\alpha_{\mathrm{E}},\alpha_{\mathrm{P}}$}
    \KwInit {$\Th_e, 
            \{ \bb_n \}_{n=1}^N, 
            \{ {\Th_{\u}^m} \}_{m=1}^M
            $ with random values;
            $ \{ \h_n \}_{n=1}^N
            $} with zeros.
        
    \While {not converged}{
        Set $\mathcal{D}_{+,h}=\{\bm{\mathcal{S}}^{(n)}_{+,h}\}_{n=1}^N = \{\bm{\mathcal{S}}^{(n)} + \{\h_n\}\}_{n=1}^N$\;
        \For{$i_{\mathrm E}=1:n_{\mathrm E}$}{
            Update PANet parameters\;
            Get intermediate features:
            $\{\bb_n = \bm{\mathrm{e}}^{(n)}\}_{n=1}^N$\;
        }
        Set $\mathcal{D}_{+,b}=\{\bm{\mathcal{S}}^{(n)}_{+,b}\}_{n=1}^N = \{\bm{\mathcal{S}}^{(n)} + \{\bb_n\}\}_{n=1}^N$\;
        \For{$i_{\mathrm P}=1:n_{\mathrm P}$}{
                Update the reconstruction parameters and common representation\;
        }
    }
    \KwOut {$\{ \Th_\mathrm{E}^m \}_{m=1}^M$ and $\{\h_{n_t}\}_{n=1}^N$;} 
    \BlankLine
    \BlankLine
    \textbf{Testing Stage}\par
    \KwIn{Incomplete testing conversation dataset: $\mathcal{D}=\{ (\bm{\mathcal{T}}^{(n)}, \bm{y}^{(n)}) \}_{n=1}^N$}
    \KwInit{Initialize $\{\Th_{u t}^m\}_{m=1}^M$ with $\{\Th_{u}^m\}_{m=1}^M$;}
        Fine-tune networks parameters\;
    Get common representations $\h^*$ of testing data set\;
    Set $\{\bm{\mathcal{T}}^{(n)}_{+,h}\}_{n=1}^N = \{\bm{\mathcal{T}}^{(n)} + \{ \h^* \}\}_{n=1}^N$\;
    Predict test emotion labels with extended dataset $\mathcal{D}^*=\{ (\bm{\mathcal{T}}^{(n)}_{+,h}, \bm{y}^{(n)}) \}_{n=1}^N$\;
    \KwOut{Predicted emotion labels $\hat{y}$.}
\end{algorithm}
\vspace{-2em}\newline

Then the extended data set $\mathcal{D}_{+,h}$ is sent to PANet model to train and update the ERC task parameters $\Th_\mathrm{E}$, getting intermediate information $\bb$ at during the conversation utterance at different time of the model. This intermediate features, containing conversation topic related(spatial) and time context(temporal) information, can reinforce the CRL by extending data set to extended one $\mathcal{D}_{+,b}$ and produce more accurate, abundant common representation $\{\h_n\}_{n=1}^N$. Finally the common representation is combined with input incomplete data set $\mathcal{D}$, producing new extended data set $\mathcal{D}_{+,h}$ and using it to train PANet model.

A big challenge lies in narrowing the gap between training and testing in multimodal representation learning, originating from their difference. We introduce a fine-tuned strategy to address this problem. The consistency of common representation from training to testing stage is ensured by optimizing the objective function $\mathcal{L}^\prime_{\mathrm{P}}$ defined in Eq.(\ref{eq:Lp_test}), getting a new set of parameters $\{\Th_{u t}^m\}_{m=1}^M$ that applied in test stage. Then test common representation can be calculated by CRL $\h^*$. Finally, this common representation is sent to the trained PANet model with test utterance (incomplete) to get predicted emotion labels.

\subsection{Party Attentive Network}
Apparently the emotion $e_t$ is mainly determined by the utterance $\u_t$. In addition, we assume that (1) the conversation scene has context by proceeding utterances; (2) every party(speaker) has its emotional context by its utterances 3) the emotion of one party is affected by the other party's emotion context.

Our PANet model is shown in Fig.\ref{fig:ep}(b). We model the conversation scene state by \textit{global state}, which is updated by the previous speaker's party-states, emotional states and current speaker's utterance, to encode shared information between all parties. It tracks party state, emotional state and utterance of all parties, using all proceeding global state jointly with utterance to get global context for party state.

Furthermore, each party is modeled by \textit{party state}, which changes as a new utterance flows in, tracking parties' dynamic representation related to the emotional state. Like global context, all proceeding party states with utterance are exploited to get party context for emotion state.
Finally, the model infers the emotional states of different parties from the corresponding party states along with other parties'. The emotional states are then sent to the classification head to get the label.

\noindent \textbf{Global State} \quad The purpose of global state  ($g\in\R^{D_\mathcal{G}}, D_\mathcal{G}$ represents dimension) is to model and track conversation scene and shared information between its parties. Current state $g_t$ focuses on current utterance's impact while proceeding global ones $g_1,\dots,g_{t-1}$ reveals the change of conversation representation, which provide meaningful context for party state. At time $t$, the global state $g_t$ updated through $\GRU_{\G}$ as follows:
\begin{equation}
    g_t=\GRU_{\G}(g_{t-1}, \u_t \oplus p^{q_{I(t-1)}}_{t-1} \oplus e^{q_{I(t-1)}}_{t-1}),
\end{equation}
where $p^{q_{I(t-1)}}_{t-1}$ is the previous speaker's party state, $e^{q_{I(t-1)}}_{t-1} $ is the corresponding emotion state and $\oplus$ means concatenation. 

\noindent \textbf{Party State} \quad PANet defines party state $p_1,p_2,\dots,$ $p_Q\in$ $\R^{D_\mathcal{P}}$ for each party in conversation scene, where $D_\mathcal{P}$ is the dimension of party state. It represents the speakers' role in a conversation and serves as bridge between global state with corresponding emotion state. Proceeding party state can also be exploited as party context attending to improve the accuracy of emotion state. 

To fully take the global state into consideration, we employ Attention Mechanism \cite{attention} to calculate global context $c^g_t=\ATTN([g_1, g_2,\dots,g_t], \u_t)$ related to current utterance:
\begin{equation}
    \begin{split}\label{eq:attn}
        \alpha_t &= {\mathrm{softmax}}(\u^T_t W_\alpha [g_1,g_2,\dots,g_t]), \\
        c_t^g &= \alpha_t^T [g_1, g_2,\dots,g_t],
    \end{split}
\end{equation}
where $\alpha_t \in \R^{t-1}$ is the attention weight which will be higher for important factors assigned for all proceeding global state related to current utterance, $c_t\in \R^{D_\mathcal{G}}$ is the weighted sum of proceeding global state by $\alpha_t$. The scheme of attention module for global is shown in Fig.\ref{fig:ep}(b).

Then party state $p_q$ is updated respectively by employing global context, current utterance and previous emotion state:
\begin{equation}
    p^q_t = \GRU_\P^q(p^q_{t-1}, \u_t \oplus c_t \oplus e^q_{t-1}), \quad q\in \{1,2,\dots,Q\},
\end{equation}
where $p^q_{t-1}$ is previous party state, $e^q_{t-1}$ is previous emotional state. Notice that party attention updates using its party and emotion state instead of the previous speaker's, which is different from global states.

\noindent \textbf{Emotion State} \quad We assume that every party in the conversation has its emotional state behind party state and is affected by other ones, and it varies under different global states. It shows how the conversation scene and other parties' state can affect party emotion. Thus we exploit all parties' state related to global state to represent party context:
\begin{equation}
        c^{p_q}_t = \ATTN([p^q_1, p^q_2,\dots, p^q_{t}], g_t),\quad q \in \{1,2,\dots,Q\},
\end{equation}
where function $\ATTN$ is defined as Eq.(\ref{eq:attn}), $p^q_t$ is the state of party $q$ at time $t$ and $c^{p_q}_t \in \R^{D_\mathcal{P}}$ is the party context of $p_q$. The scheme of attention module for party context is shown in Fig.\ref{fig:ep}(b).

All the party context $\{c^{p_1}_t,c^{p_2}_t,\dots,c^{p_Q}_t\}$ and current global state $g_t$ are then used to update emotion state by
\begin{equation}
    \begin{split}
        e^q_t = \GRU_\E(e^q_{t-1}, c^{p_1}_t\oplus c^{p_2}_t \oplus \dots c^{p_Q}_t \oplus g_t), \\ 
        \quad q \in \{1,2,\dots,Q\},
    \end{split}
\end{equation}
where $e^q_t \in \R^{\mathcal{E}}$.

\noindent \textbf{Emotion Classification Head}\quad
We employ a simple perceptron $\mathcal{C}$ to classify the speaker's emotion state as final emotion label:
\begin{equation}
    \hat{y}_t = \arg\max \mathcal{C}(e^{I(t)}_t),
\end{equation}
where $\hat{y}_t$ is the predicted emotion label.

Finally, we employ categorical cross-entropy along with L2-regularization as the objective function:
\begin{equation}
    \mathcal{L}_\mathrm{E}(\bm{\x}, \bm{y}) = -\frac{1}{T}
    \sum_{t=1}^T{
    \log{\mathcal{P}_{t}[y_{t}]+\lambda_\mathrm{E}\Vert \Th_\mathrm{E} \Vert_2}
    }
\end{equation}
where $\bm{\x}, \bm{y}$ is the time-sequence data with labels, $T$ is the total number of utterance in a conversation, $\mathcal{P}_{t}$ is the distribution of emotion labels for utterance $\u_t$, $y_t$ is the ground truth, $\lambda_\mathrm{E}$ is the L-2 weight and $\Th_\mathrm{E}$ represents the parameters of PANet model.

\subsection{Common Representation Learning}
Enlightened by \cite{partial} and take the assumptions in section \ref{sec:Pdefinition}, we use CPM-Nets to learn common representation. 
Model $p(\u_i|\h)$ as follows:
\begin{equation}
    p(\u^m|\h) \propto e^{-l_r(\u^m,f_m(\h;\Th_u^m))},\quad m\in\{1,2,\dots,M\},
\end{equation}
where $\Th_u^m$ are the parameters responsible for the mapping $f_i(\cdot)$ from common representation $\h$ to utterance modality $\u^m$, with $l_r(\cdot)$ being the reconstruction loss function.

Thus, the log-likelihood function can be written as
\begin{equation}
    \begin{split}
    \mathcal{L}\left(\{\h_n\}_{t=1}^T,\Th_u^1,\dots,\Th_u^M,\Th_y)\right) =\sum_{t=1}^T{\ln{p(y_t,\cS_t|\h_t)}} \\
    \propto -\left( l_c(y_t,g(\h_t|\Th_y))+ \sum_{t=1}^T{\sum_{m=1}^M{l_r(\u^m,f_m(\h;\Th_u^m))}}\right).
    \end{split}
\end{equation}

The common representation $\h_t$ is used to reconstruct data $\u^m$ by the loss $l_r(\u^m,f_m(\h;\Th_u^m))$, which is balanced by the classification loss $l_c(y_n,g(\h_t|\Th_y))$ encourage the alignment with the ground truth.

For classification task, we define reconstruction loss function as
\begin{equation}
    \begin{split}\label{eq:reconstruction}
        l_r(\u^m,f_m(\h;\Th_u^m)) = \Vert \u^m-f_m(\h;\Th_u^m)) \Vert ^2 , \\
        \quad m\in\{1,2,\dots,M\},
    \end{split}
\end{equation}
where the mapping direction is from common representation to different modalities, making sure all the samples (regardless of their missing patterns) comparable.

Accordingly, the classification loss is defined as
\begin{equation}
    \begin{split}
        l_c(y_t,\hat{y}_t) = \max(0, \Delta(y_t,\hat{y})&+\mathbb{E}_{\h\sim\mathcal{H}(y)}{F(\h,\h_t)} \\
        &- \mathbb{E}_{\h\sim\mathcal{H}(y_t)}{F(\h,\h_t)}
        ), 
    \end{split}
\end{equation}
where $\hat{y}=\arg_{y\in\mathcal{Y}}\max{\mathbb{E}_{\h\sim\mathcal{H}(y)}{F(\h,\h_t)}}$ is the predicted label and $\mathcal{H}(y)$ is the set of common representation for $y$.The loss $\Delta(y_t,\hat{y})$ is 0 if $y_t=\hat{y}$ else 1.


To better employ observed data and to promote data imputation, we introduce adversarial strategy in our algorithm. Learning common representation is guided simultaneously by observed data, unobserved data and labels. The adversarial loss is calculated by wasserstein GAN(WGAN) \cite{wgan} as:
\begin{equation}
\begin{split}
    \mathcal{L}_{\mathrm{A}} = \sum_{m=1}^M\sum_{i=1}^I
    (1-s_{m})[D_m(\hat{\u}^m; \Th_{d}^m)\\
    -D_m(\u_i^{m};\Th_{d}^m)+\lambda_{\mathrm{G}}\mathcal{P}_{\mathrm{G}}],
\end{split}
\end{equation}
where $\hat{\u}^m=f_m(\h_n;\Th_{\u}^m)$ is the generated missing view data, $\mathcal{P}_{\mathrm{G}}=(\Vert\nabla_{\hat{\u}^m}D_m(\hat{\u}^m)\Vert_2-1)^2$ is the gradient penalty in WGAN and $\lambda_{\mathrm{G}}$ is the balanced parameter. $(1-s_{m})$ indicates whether or not the sample at scene $n$ and modality $m$ should be computed. $\{\u_i^m\}_i^I$ are the observed data, with the total number of samples being $I$. The discriminator $D_m$ of $m$-th view defined by parameter $\Th_{d}^m$ determinate the imputation data between the observed ones.

Finally, the overall loss function is shown as:
\begin{equation}
    \begin{split}
        \mathcal{L}_{\mathrm{P}}(\bm{\cS}, \bm{y}) &= \lambda_{\mathrm{R}}\mathcal{L}_{\mathrm{R}}(\bm{\cS}, \h) + \lambda_{\mathrm{C}}\mathcal{L}_{\mathrm{C}}(\bm{y}, \h) +
        \lambda_{\mathrm{A}}\mathcal{L}_{\mathrm{A}}(\bm{\cS}) \\
        \mathcal{L}_{\mathrm{R}}(\bm{\cS}, \h) &= \frac{1}{T}\sum_{t=1}^T\sum_{m=1}^M{l_r(\u^m,f_m(\h;\Th_u^m))} \\
        \mathcal{L}_{\mathrm{C}}(\bm{y}, \h) &= \frac{1}{T}\sum_{t=1}^T{l_c(y_t,y,\h_t)}, \\
    \end{split}
\end{equation}
where $\bm\cS, \bm{y}$ is the incomplete input data with labels and $\h$ is the common representation learned by reconstruction function $f_m$. The hyper-parameters $\lambda_\mathrm{R}, \lambda_{\mathrm{C}}, \lambda_{\mathrm{A}}$ are used to adjust reconstruction loss, classification loss and adversarial loss respectively.

At testing stage, we employ fine-tune strategy to ensure the consistency by adjusting the loss function defined as Eq.(\ref{eq:reconstruction}) to narrow the gap between training set and testing set:
\begin{equation}\label{eq:Lp_test}
    \mathcal{L}^\prime_{\mathrm{P}}(\bm\cS^\prime) = {l_r(\bm\cS^\prime,f^\prime_m(\h;\Th_{u t}^m))},\quad m\in\{1,2,\dots M\},
\end{equation}
where $\Th_{u t}^m$ is the parameters that control the mapping from test input to its common representation, and $\bm\cS$ represents the incomplete testing dataset.

\section{Experiment Setup}\label{sec:result}
\subsection{Datasets}
We test our model and algorithm on the ERC datasets: IEMOCAP \cite{iemocap}, and MELD \cite{meld}. IEMOCAP is a two-party dataset, while MELD is a multiple-party one. We split both the datasets into train/val and test samples approximately by the ratio 80/20, guaranteeing that they do not share any party. The detailed conversations' and utterances' distribution of train/val/test samples are shown in Table \ref{tbl:dataset}.

\vspace{-1mm}
\begin{table}[htbp]\label{tbl:dataset}
\resizebox{0.47\textwidth}{!}{
\begin{tabular}{lcccc}
\hline
Dataset & \multicolumn{1}{l}{partition} & \multicolumn{1}{l}{\#conversations} & \multicolumn{1}{l}{\#utterances} & \multicolumn{1}{l}{\#classes} \\ \hline
\multicolumn{1}{c}{\multirow{2}{*}{IEMOCAP}} & train/val                     & 132 & 5810 & \multirow{2}{*}{7} \\
\multicolumn{1}{c}{} & test & 31 & 1623 & \\ \hline
\multicolumn{1}{l}{\multirow{2}{*}{MELD}} & train/val & 1153 & 11098 & \multirow{2}{*}{3\&7} \\
\multicolumn{1}{l}{} & test & 280 & 2610 & \\ \hline
\end{tabular}}
\vspace{2mm}
\caption{Statistics of two datasets.In MELD, the annotation contains both sentiments(\#3) and emotions(\#7).} 
\vspace{-2mm}
\end{table}

\noindent\textbf{IEMOCAP}\quad This dataset is acted, multimodal and multi-speaker containing approximately 12 hours of dyadic sessions with markers on the face, head, and hands. The actors performed selected emotional scripts and also improvised hypothetical scenarios designed to elicit specific types of emotions (\textit{happiness, anger, sadness, frustration, neutral state}).

\noindent\textbf{MELD}\quad Multimodal EmotionLines Dataset has been created by enhancing and extending the EmotionLines dataset. MELD contains the same conversation instances available in EmotionLines, but it also encompasses audio and visual modality along with the text. MELD has more than 1400 conversations and 13000 utterances from the Friends TV series. Multiple speakers participated in the conversations. Each utterance in the conversation has been labeled by any of these seven emotions (\textit{happiness, anger, sadness, disgust, surprise, fear, neutral}).

\subsection{Feature Extraction}\label{sec:feature}
\noindent\textbf{Textual Feature}\quad
We employ convolutional neural networks (CNN) for textual feature extraction. Following \cite{textfeat}, we obtain n-gram features from each utterance using three distinct convolution filters of sizes 3,4 and 5 respectively, each having 50 feature maps. Outputs are then subjected to max-pooling and rectified linear unit (ReLU) activation. These activations are concatenated and fed to a 100-dimensional dense layer, regarded as the textual utterance representation. Then the whole network is trained with the emotion labels.

\noindent\textbf{Audio and Visual Feature}\quad
We follow the identical procedure with \cite{dialoguernn}, employing 3D-CNN and openSMILE \cite{opensmile} to extract audio and visual feature \cite{3dcnn}, respectively.

\subsection{Baselines}
\noindent\textbf{bc-LSTM}
Biredectional LSTM \cite{bclstm} is used to capture thecontext from the surrounding utterances to generate context-aware utterance representation. However, this model does not differentiate among the speakers.

\noindent\textbf{DialogueRNN} models the emotion of utterances in a conversation with speaker, context and emotion information from neighbour utterances\cite{dialoguernn}. These factors are modeled using three separate GRU networks to keep track of the individual speaker states.

\noindent\textbf{ICON} uses two GRU networks to learn the utterance representations for conversations between two participants. The output of the two speakers' GRUs is then connected using another GRU that helps perform explicit inter-speaker modeling.

\noindent\textbf{KET} or Knowledge enriched transformers dynamically leverages external commonsense knowledge using hierarchical self-attention and context-aware graph attention.

\noindent\textbf{MMER} is composed of several sub-modules: three feature extractors, three unimodal classifiers, three bimodal classifiers and one trimodal classifier\cite{MMER}. It leverages a multimodal multitask approach to tackle the problem of missing modality at test time and training time.

\noindent\textbf{SMIL} uses building features reconstruction network to solve the problem of missing modality. It uses a Bayesian neural network to assess the data uncertainty as feature regularization to overcome model and data bias produced by the feature reconstruction network\cite{SMIL}.

\noindent\textbf{SMVAE} essentially performs infinite imputations by integrating out the missing data\cite{SMVAE}. It first develops a multi-view variational autoencoder model for fusing multi-modality emotional data and builds a semi-supervised emotion recognition algorithm.

\subsection{Network Architectures}
We employ bi-directional GRU \cite{birnn} cells in our PANet model, where global state dimension $D_\mathcal{G}$, party state dimension $D_\mathcal{P}$, emotional state dimension $D_\mathcal{E}$ being $512,512,256$ respectively. For emotion classification head, a simple perceptron composed of 2-layer linear layers attached with ReLU activation is used.
For CPM-Nets, we employ the fully connected networks with batch normalization and LeakyReLU for all generators $f_m$. Similarly, fully connected networks with LeakyReLU activation structures are employed for discriminators $D_m$. L2 regularization is used with the value of the trade-off parameter being 0.001 for $f_m$ and $D_m$.

\subsection{Parameter Settings}
At training state, we use Adam optimizer \cite{adam} for all parameter update, where $\beta_1=0.5,\beta_2=0.999$. The learning rate for PANet model $\alpha_\mathrm{E}$ is $1\text{e-}4$ with the L2 regularization being $1\text{e-}5$. And learning rates for the generators and discriminators $\alpha_\mathrm{E}$ in CPM-Nets are all $1\text{e-}3$. The trade-off parameters $\lambda_{\mathrm{R}},\lambda_{\mathrm{P}},\lambda_{\mathrm{A}}$ of reconstruction, classification and adversarial losses in CRL problem are $1,10,10$ respectively. Specifically, the parameter for gradient penalty $\lambda_{\mathrm{G}}$ in WGAN is default as $1$. And we train generators after discriminators update 2 times in order to guarantee the stability in GAN structure \cite{gantrain}. In each M2R2 epoch, we train PANet model $n_\mathrm{E}$ times and CPM-Nets for $n_\mathrm{P}$ times.

\section{Results and Analysis}
\subsection{PANet Compared with Baselines} 
\begin{figure*}[htbp]
    \centering
    \subfigure[IEMOCAP]{\label{fig:acc_iemocap}
    	\centering
    	\begin{minipage}{0.45\linewidth}
    		\centering
    		\includegraphics[width=\linewidth]{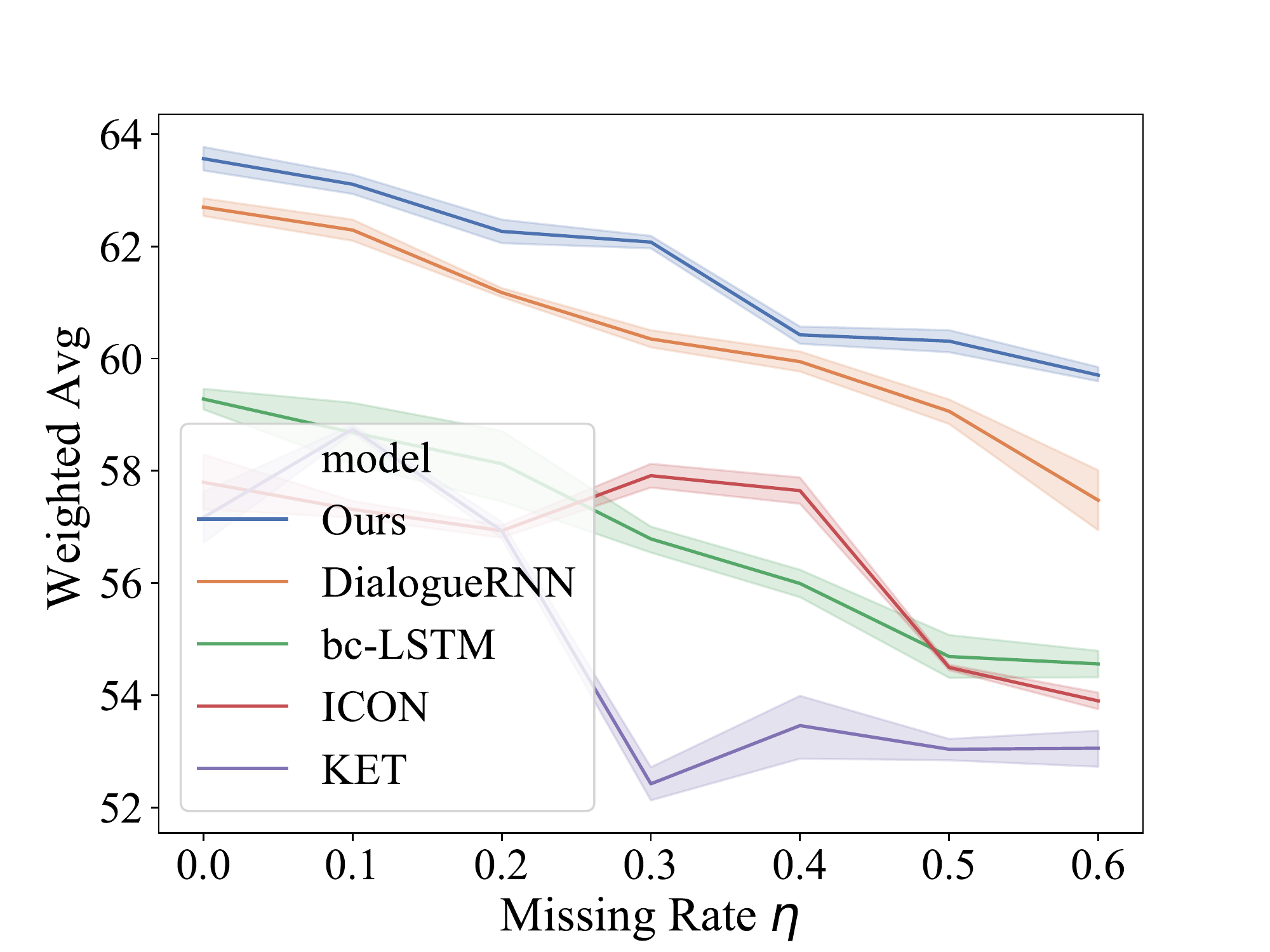}
    	\end{minipage}
    	\begin{minipage}{0.45\linewidth}
    		\centering
    		\includegraphics[width=\linewidth]{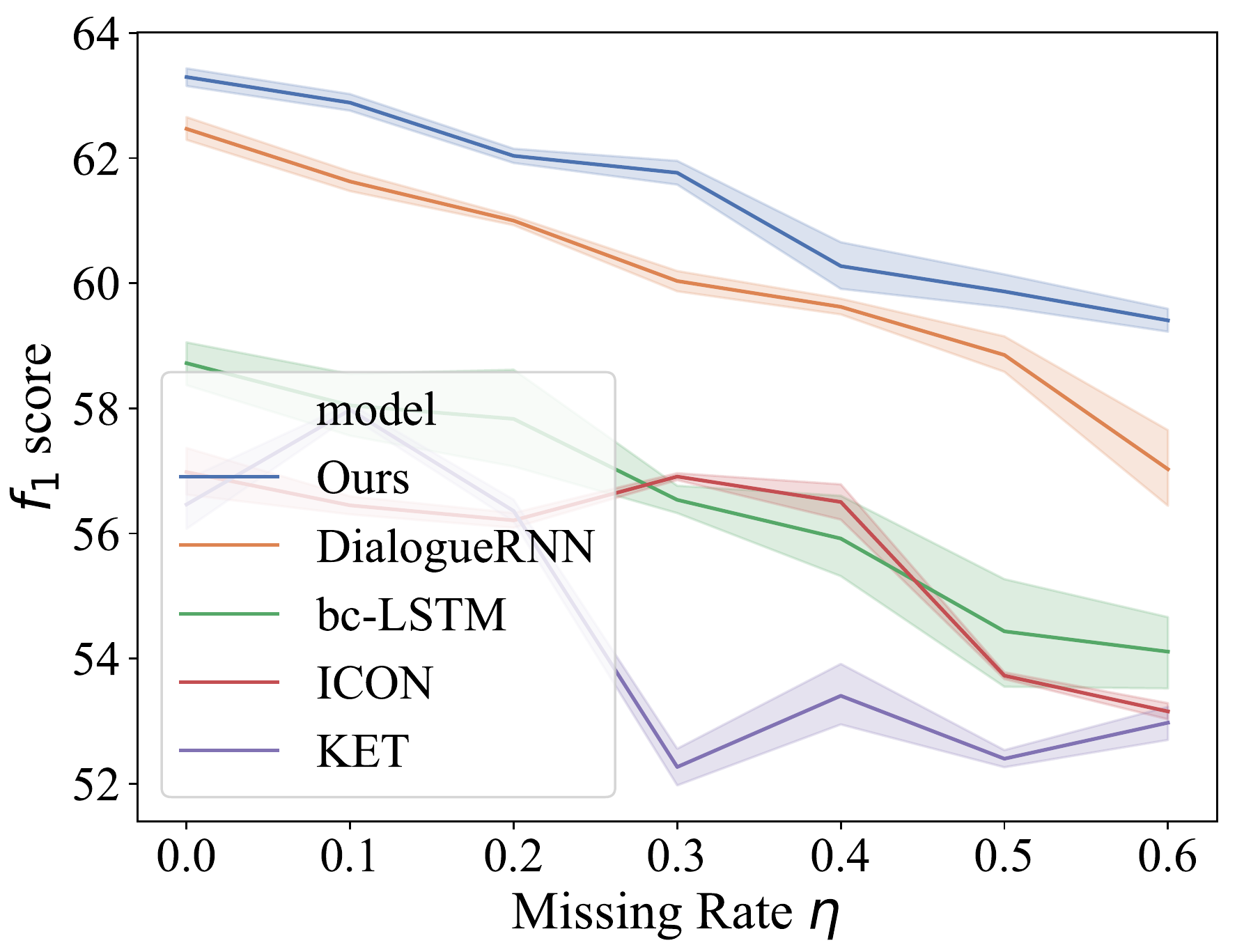}
    	\end{minipage}
    }
    \subfigure[MELD]{\label{fig:f1_meld}
    	\begin{minipage}{0.45\linewidth}
    		\centering
    		\includegraphics[width=\linewidth]{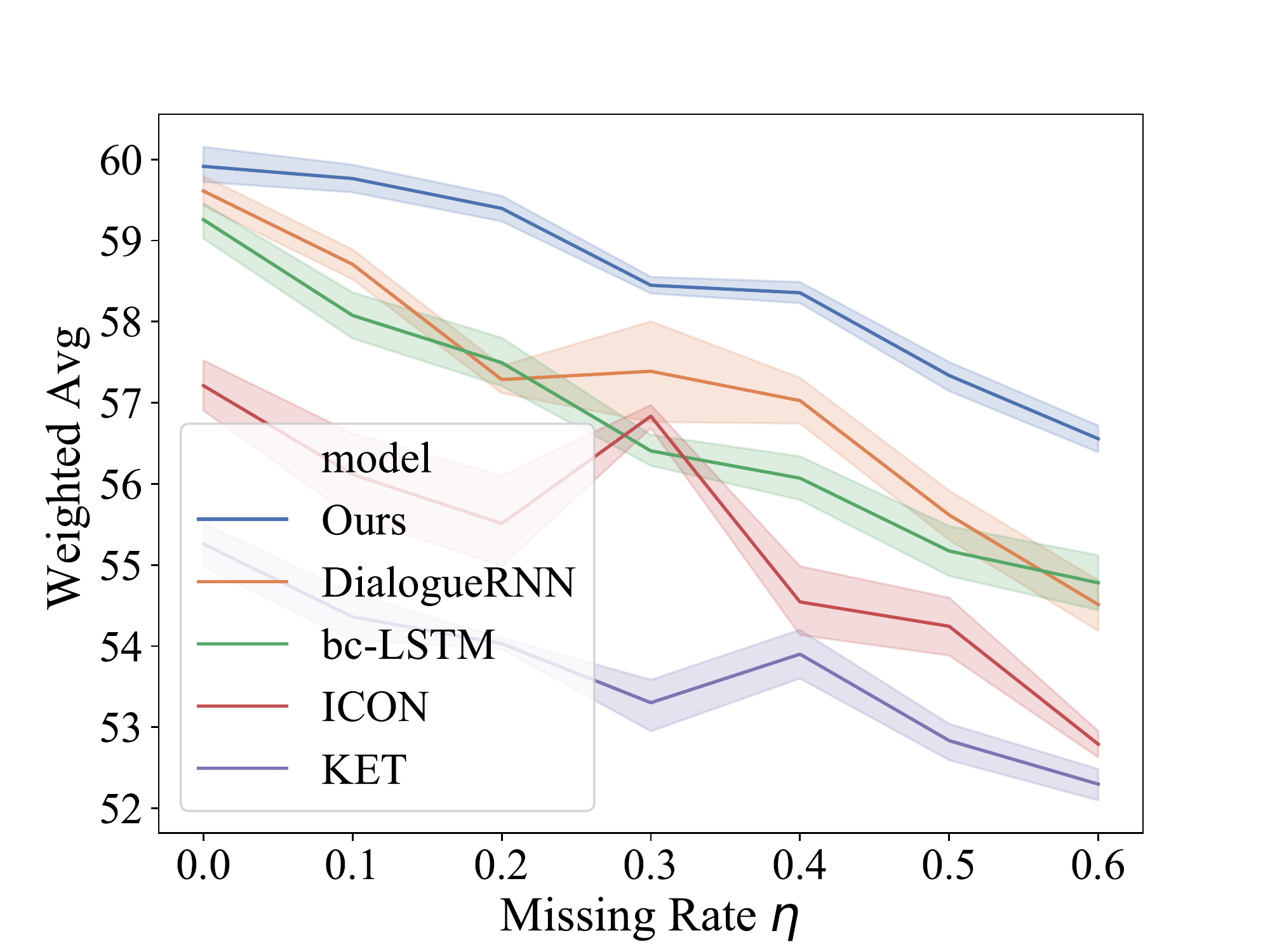}
    	\end{minipage}
    	\begin{minipage}{0.45\linewidth}
    		\centering
    		\includegraphics[width=\linewidth]{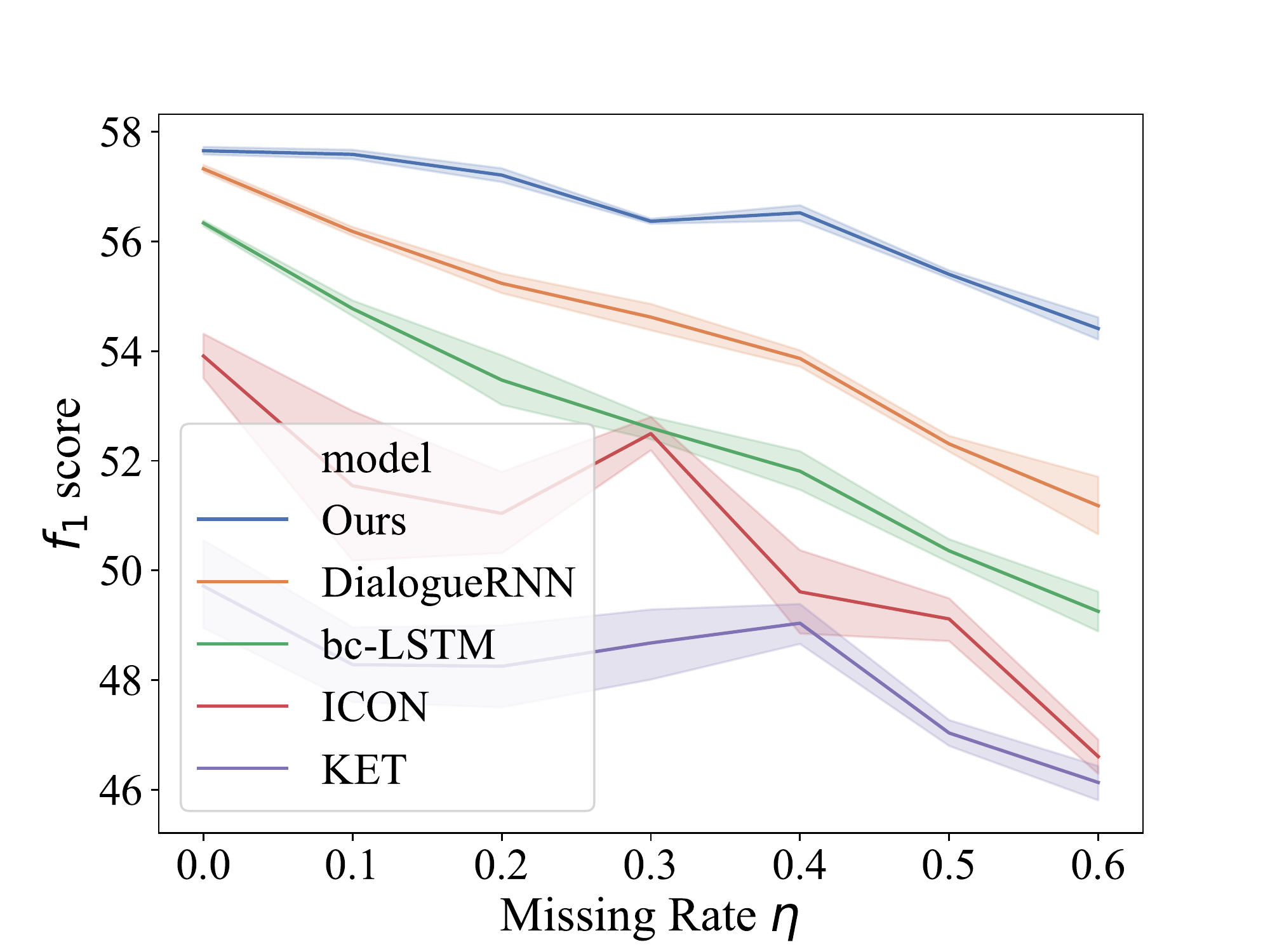}
    	\end{minipage}
    }
	\caption{Results of different emotion recognition models with missing rate.}
	\label{fig:matric_datasets}
\end{figure*}
Firstly we test our PANet model and baselines at different missing rate defined as $\eta={\sum_t\sum_m{S_t}}/{\sum_t\sum_m{\u_t^m}}$ in section \ref{sec:Pdefinition} ranging from $0.0$ to $0.6$ by step $0.1$ repeatedly 5 times. To guarantee that no blank utterance is sent to the model, every row of the missing matrix must have one element, where zeros vector is used here for the absent modalities. Additional hidden representations are not used here. 

Comparison results of IEMOCAP\&MELD datasets are shown in Fig \ref{fig:matric_datasets}, where there are the following observations: 1) Our PANet model performs better at both accuracy and F1-score than baselines when the dataset is complete; 2) As missing rate $\eta$ increases, the degradation of our model is much faint than baselines, especially on MELD dataset. For example, the relative dropping accuracy of PANet at MELD dataset from $\eta=0$ to $\eta=0.3$ is $1.55\%$, while the relative dropping accuracy of the second-best DialogueRNN is $2.96\%$; 3) Our model is more robust to the modalities missing rate than other baselines. For instance, the variance of our PANet model related to the missing rate $\eta$ on the MELD dataset is $2.364$ compared with $3.54099$ of DialogueRNN.

\begin{table*}[htbp]
\centering
    \subtable[IEMOCAP]{\centering\label{tbl:iemocap}
    \resizebox{0.8\linewidth}{!}{
    \begin{tabular}{ccccccccccccccc}
\hline
\multirow{2}{*}{Methods} & \multicolumn{2}{c}{Happy}       & \multicolumn{2}{c}{Sad}         & \multicolumn{2}{c}{Neutral}     & \multicolumn{2}{c}{Angry}       & \multicolumn{2}{c}{Excited}     & \multicolumn{2}{c}{Frustrated}  & \multicolumn{2}{c}{Average(w)}  \\ \cline{2-15} 
 & Acc & F1 & Acc & F1 & Acc & F1 Acc & F1 & Acc & F1 & Acc & F1 & Acc & F1 \\ \cline{2-15} 
bc-LSTM                  & 38.74          & 24.24          & 69.34          & 62.35          & 48.72          & 52.35          & 54.43          & 56.33          & 56.32          & 70.33          & 55.38          & 52.37          & 54.28          & 52.99          \\
ICON                     & 51.16          & 20.09          & 73.73          & 60.80          & 40.97          & 46.05          & \textbf{67.87} & 56.87          & 70.48          & 66.11          & 50.07          & 56.07          & 54.49          & 53.73          \\
KET                      & \textbf{56.82} & 23.12          & 76.18          & 65.44          & 42.44          & 48.66          & 65.90          & 58.76          & 67.65          & 62.20          & 51.62          & 57.41          & 55.12          & 54.53          \\
DialogueRNN              & 52.89          & 18.65          & \textbf{79.53} & 59.24          & 39.01          & 46.31          & 58.15          & 53.97          & \textbf{71.34} & 63.20          & 50.89          & 57.73          & 57.11          & 52.40          \\ \hline
SMVAE                    & 39.34          & 25.27          & 71.42          & 65.63          & 49.24          & 53.24          & 55.24          & 56.33          & 58.42          & 73.24          & 54.24          & 53.22          & 55.33          & 54.49          \\
MMER                     & 41.74          & 26.23          & 73.24          & 67.64          & 54.03          & 55.24          & 57.26          & 57.33          & 60.24          & 74.27          & 56.39          & 53.24          & 56.83          & 55.66          \\
SMIL                     & 40.35          & 27.50          & 76.82          & 66.78          & 53.03          & 55.10          & 58.73          & \textbf{59.53} & 63.87          & \textbf{78.13} & 57.39          & 55.85          & 58.68          & 57.42          \\ \hline
Ours                     & 41.35          & \textbf{32.84} & 77.82          & \textbf{71.42} & \textbf{54.72} & \textbf{56.72} & 59.73          & 59.49          & 64.87          & 70.80          & \textbf{57.46} & \textbf{58.51} & \textbf{59.64} & \textbf{59.17} \\ \hline
\end{tabular}
    }}
     
    \subtable[MELD]{\centering\label{tbl:meld}  
    \resizebox{0.85\linewidth}{!}{
    \begin{tabular}{ccccccccccccccccc}
\hline
\multirow{2}{*}{Methods} & \multicolumn{2}{c}{Happy}       & \multicolumn{2}{c}{Sad}         & \multicolumn{2}{c}{Neutral}     & \multicolumn{2}{c}{Angry}       & \multicolumn{2}{c}{Surprise}    & \multicolumn{2}{c}{Fear} & \multicolumn{2}{c}{Disgust} & \multicolumn{2}{c}{Average(w)}  \\ \cline{2-17} 
 & Acc & F1 & Acc & F1 & Acc & F1 & Acc & F1 & Acc & F1 & Acc & F1 & Acc & F1 & Acc & F1 \\ \hline
bc-LSTM                  & 43.65          & 35.27          & 24.24          & 11.24          & 62.38          & 74.27          & 35.42          & 39.46          & 36.35          & 36.35          & -           & -          & -             & -           & 51.24          & 49.32          \\
ICON                     & 37.70          & 36.94          & 0.00           & 0.00           & 62.48          & 72.03          & 36.43          & 36.21          & 24.05          & 17.64          & -           & -          & -             & -           & 52.83          & 47.04          \\
KET                      & 45.13          & 37.24          & 24.33          & 12.37          & 63.37          & 76.24          & 34.24          & 40.83          & 38.37          & \textbf{38.46} & -           & -          & -             & -           & 53.72          & 51.01          \\
DialogueRNN              & 45.68          & 39.43          & 29.16          & 13.01          & 62.25          & 71.37          & 36.52          & 38.67          & 40.02          & 34.92          & -           & -          & -             & -           & 54.48          & 50.04          \\ \hline
SMVAE                    & 42.18          & 42.18          & 14.58          & 1.80           & 63.46          & 72.34          & 38.22          & 35.68          & 30.66          & 27.40          & -           & -          & -             & -           & 54.24          & 49.12          \\
MMER                     & \textbf{46.24} & 38.24          & 25.38          & \textbf{13.24} & 66.24          & 78.32          & 34.18          & 40.24          & 38.37          & 38.37          & -           & -          & -             & -           & 54.24          & 51.68          \\
SMIL                     & 45.68          & 39.60          & 27.86          & 12.88          & 63.72          & 73.09          & 37.25          & \textbf{41.74} & \textbf{40.16} & 37.22          & -           & -          & -             & -           & 55.23          & 52.32          \\ \hline
Ours                     & 44.86          & \textbf{42.31} & \textbf{30.22} & 3.55           & \textbf{66.63} & \textbf{80.14} & \textbf{38.37} & 36.24          & 37.07          & 32.98          & -           & -          & -             & -           & \textbf{55.83} & \textbf{53.31} \\ \hline
\end{tabular}
    }}
    \caption{Detailed result of different emotion recognition models at missing rate $\eta=0.5$.}
    \label{tbl:datasets}
\end{table*}

\begin{figure}[htbp]
\begin{center}
\includegraphics[width=0.8\linewidth]{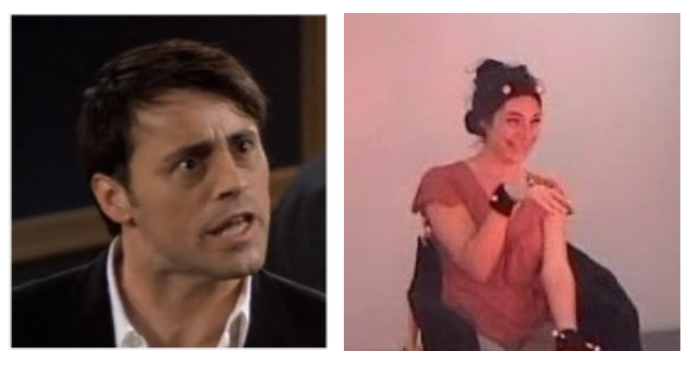}
\end{center}
   \caption{Visual/Facial modality comparison of MELD and IEMOCAP, where MELD is more expressive and vivid.}
\label{fig:compare}
\end{figure}

We can also observe that our algorithm performs better on MELD dataset than IEMOCAP regardless the accuracy or F1-score metric, the possible explanation is that MELD is a film cut of TV-shows, where the visual and audio expression is much more clear, exaggerative and severely changes than IEMOCAP as shown in Fig.\ref{fig:compare}.

\subsection{M2R2 Compared with Baselines}
We also test the detailed emotion prediction results of all classes on both datasets under the missing rate $\eta=0.5$. As Table.\ref{tbl:meld}) shows, our model performs better than baselines on both IEMOCAP and MELD datasets at missing rate $\eta=0.5$, where metrics are accuracy and F1 score. Significantly, the M2R2 framework can detect Neutral and Frustrated emotion more accurately, gaining far more accuracy and F1 socre than the second-best DialogueRNN model. We presumed that the exploit of party attention related to the global state provides more hidden emotion representation for the speaker, which neutralizes the potential opposite emotions from all parties in a conversation.

In further discussion, we test the accuracy and F1 score of both IEMOCAP and MELD datasets with other multimodal data reconstruction algorithms. The comparison is in Table \ref{tbl:datasets} as shown that our M2R2 framework achieves the highest accuracy and F1 score on all emotion classes of both datasets, where accuracy is about 1\% better than the second-best SMIL algorithm on both datasets.

\subsection{Case Study}
\noindent\textbf{Context Attention}\quad We test our global and party attention mechanism on IEMOCAP dataset when missing rate $\eta=0.2$, where the global and different parties' attention weight $\alpha_{\mathrm{G}}, \alpha_{\mathrm{P_1}}, \alpha_{\mathrm{P_2}}$ at scene time turn 7 is shown in Fig.\ref{fig:attn_case}. Observations can be summarized as follwing: 
\begin{figure}[htbp]
\begin{center}
\includegraphics[width=\linewidth]{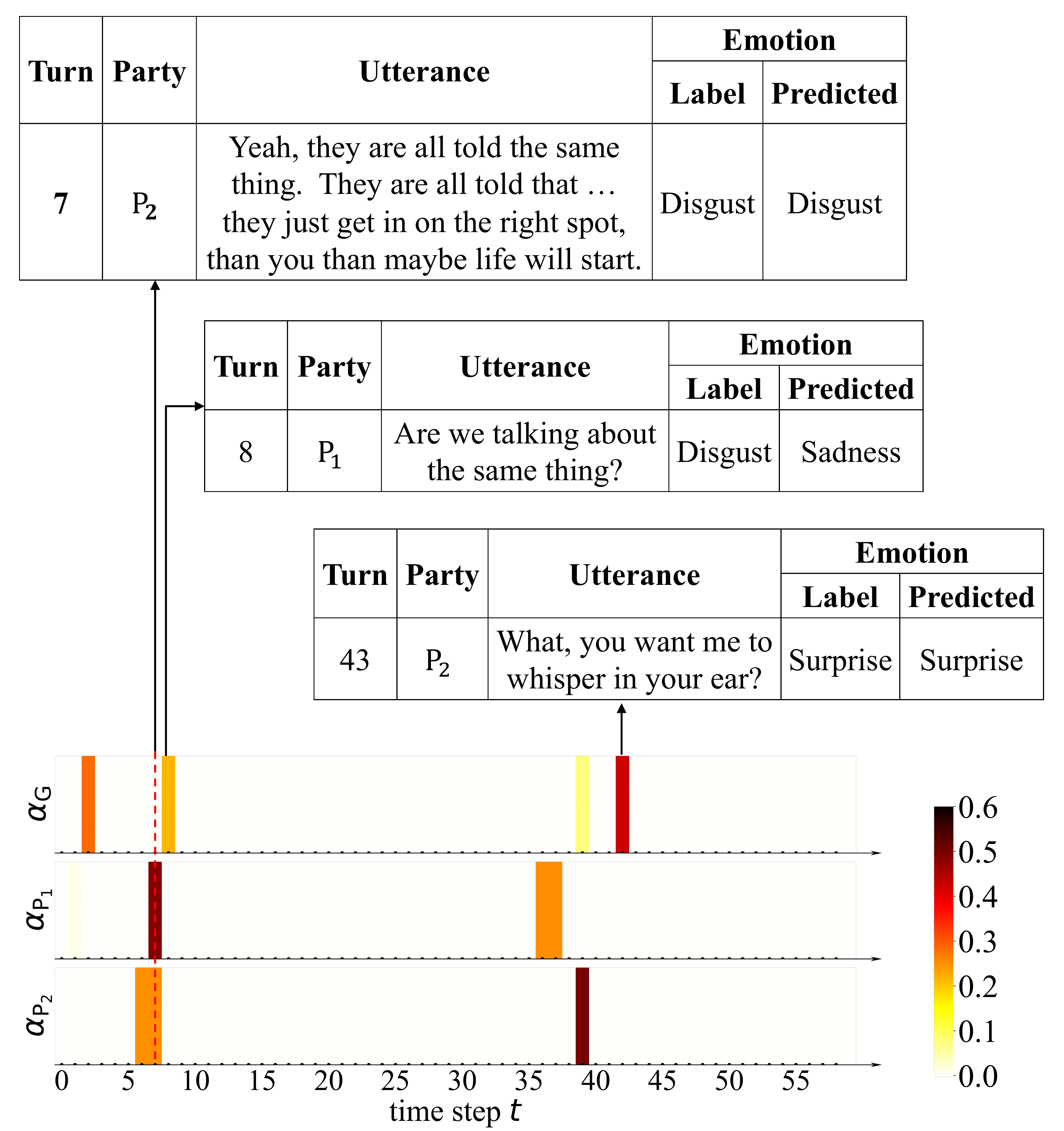}
\end{center}
    \vspace{1em}
    \caption{Global and different parties' attention weight related to all conversation utterances at scene turn 7. Note that the turn number starts from 0.}
    \label{fig:attn_case}
    \vspace{-1em}
\end{figure}

\begin{enumerate}[itemsep=5pt, topsep=5pt, parsep=0pt, itemsep=5pt]
    \item [(1)] Dependency on utterance at different distances. The attention weight of all conversation turns related to chosen turns shows that neighboring and distant utterances can influence the current emotional state. For example, the emotion at turn 7, where the model correctly predicts 'Disgust', is influenced by the nearby utterance at turn 2 and the distant utterance at turn 43. A similar dependency trend is also shared by party attention mechanism such as the nearby utterance at turn 6 and the distant utterance at turn 39 both influence the current emotion prediction at turn 7.
    \item [(2)] Dependency on future utterances. The figure also reveals that future utterances can influence the proceeding emotion greatly. In this discussion, we can observe that the turn 7 contributes almost half to proceeding turn 7 of global attention $\alpha_\mathrm{G}$ and similarly, turn 36, 37 and turn 39 have a significant influence on their proceeding turn 7 of $\alpha_{\mathrm{P_1}}, \alpha_{\mathrm{P_2}}$ respectively.
    \item [(3)] Dependency on different parties. Like common sense in daily life, one speaker's emotional state changes from the speaker's internal shift and other speakers' influence. For example, both turn 8 and 43 strongly attend to the emotion prediction of turn 7 while turn 8 is uttered by party $q_{I(8)}$ and turn 43 belongs to the other party.
\end{enumerate}

\begin{figure}[htbp]
\begin{center}
\includegraphics[width=0.95\linewidth]{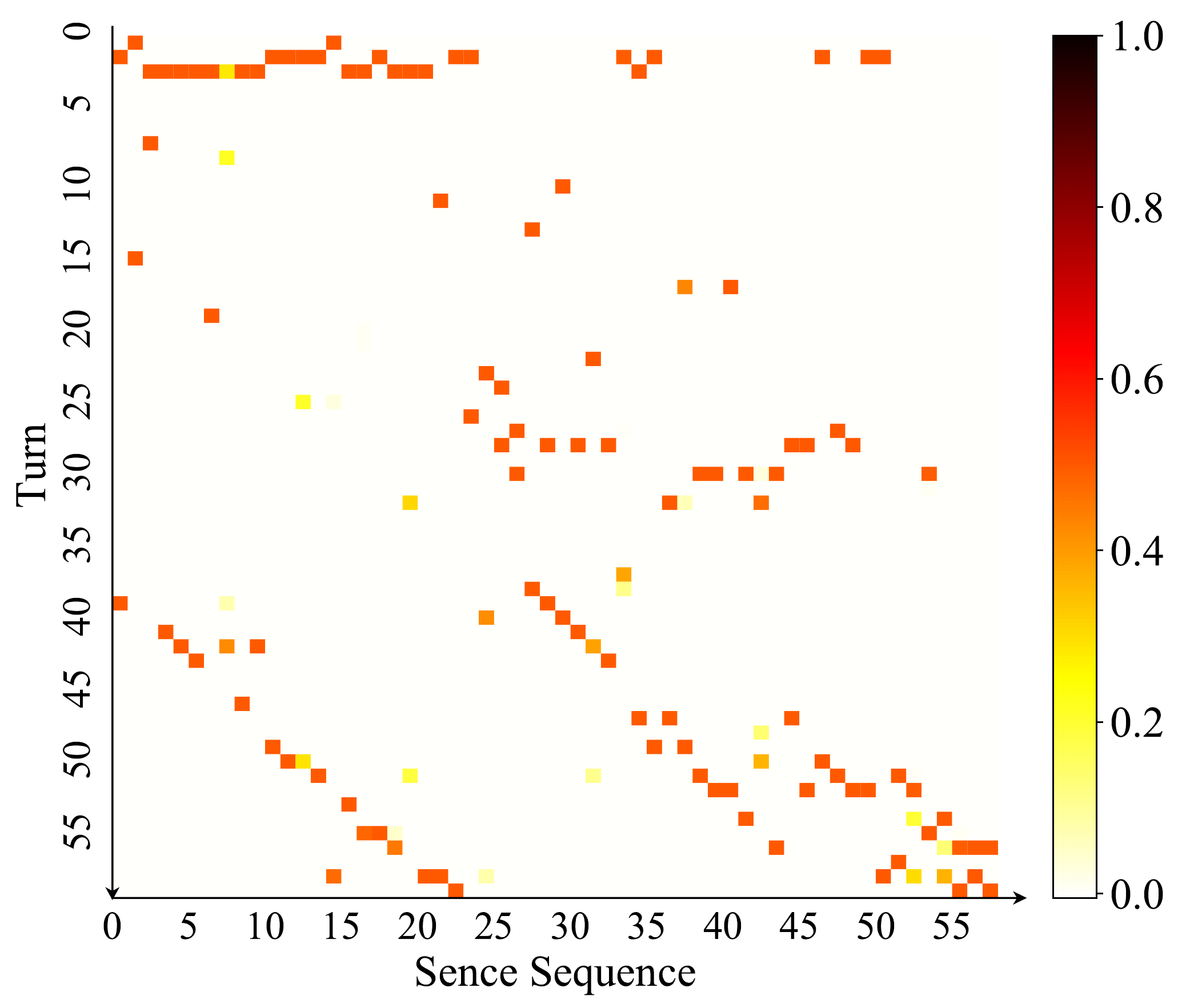}
\end{center}
    \caption{Global attention weight of all turns in one test conversation scene at IEMOCAP dataset.}
    \label{fig:global}
\end{figure}

To avoid the exception error in analysis, we test the global attention weight at all turns in a conversation scene of the IEMOCAP dataset. The Fig.\ref{fig:global} shows the result. As can be observed, early utterances attend to early scene emotion prediction more strongly than other times and accordingly, late utterances attend more strongly to late scene emotion prediction. This trend can be explained by sequential neural networks' memory capacity and time.   

\begin{figure}[htbp]
\begin{center}
\includegraphics[width=0.9\linewidth]{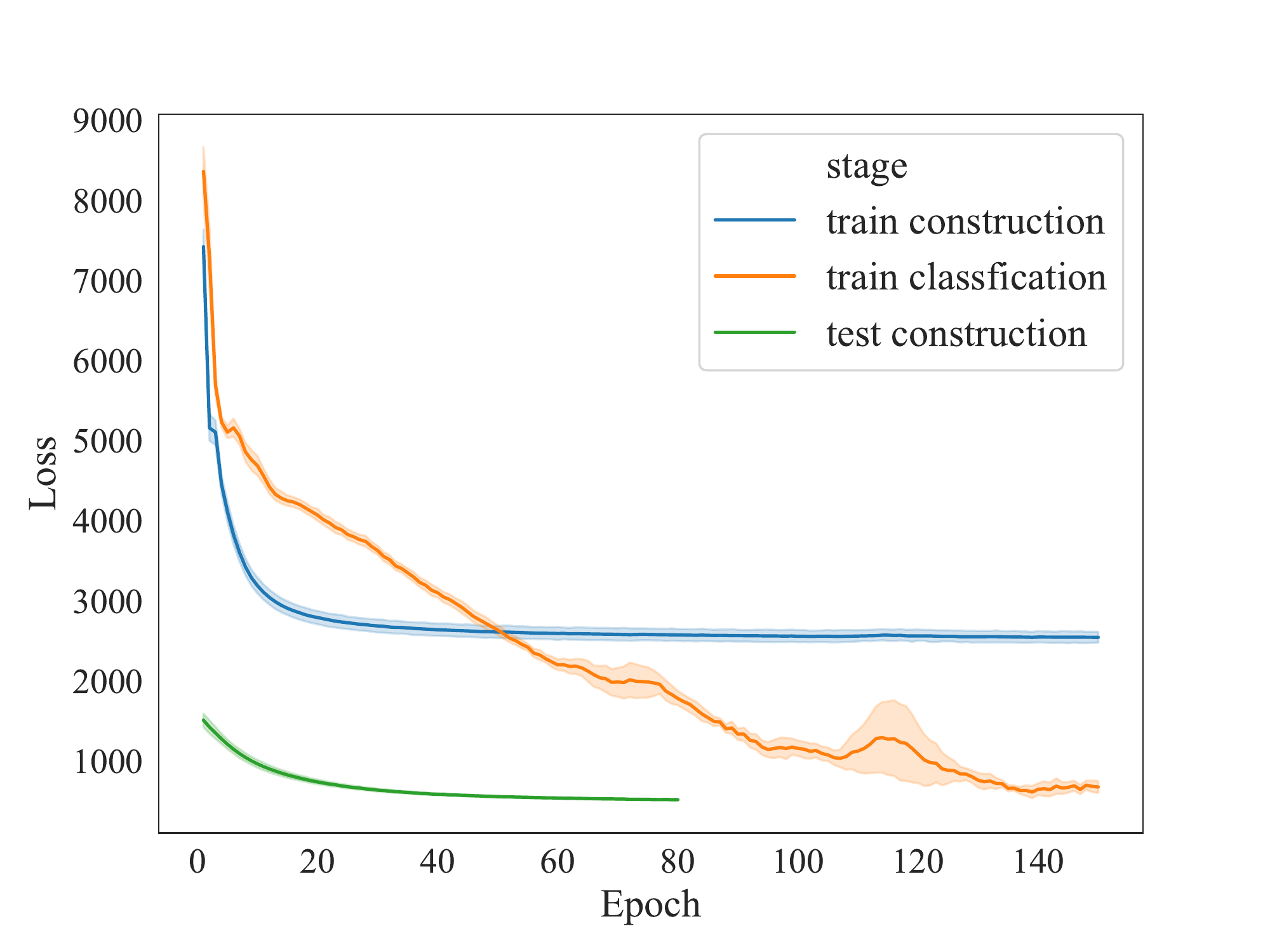}
\end{center}
    \caption{Reconstruction and classification loss at training and testing stages with epoch.}
    \label{fig:pml_loss}
\end{figure}

\noindent\textbf{CRL Convergence}\quad We show the loss curve of reconstruction and classification networks at the training and testing stage as shown in Fig.\ref{fig:pml_loss}, where the dataset used is MELD and repeated 5 times. Observations can be made that reconstruction loss converges faster and more stable than classification loss, and testing reconstruction loss has a slower convergence speed.

\begin{figure}[htbp]
\begin{center}
\includegraphics[width=\linewidth]{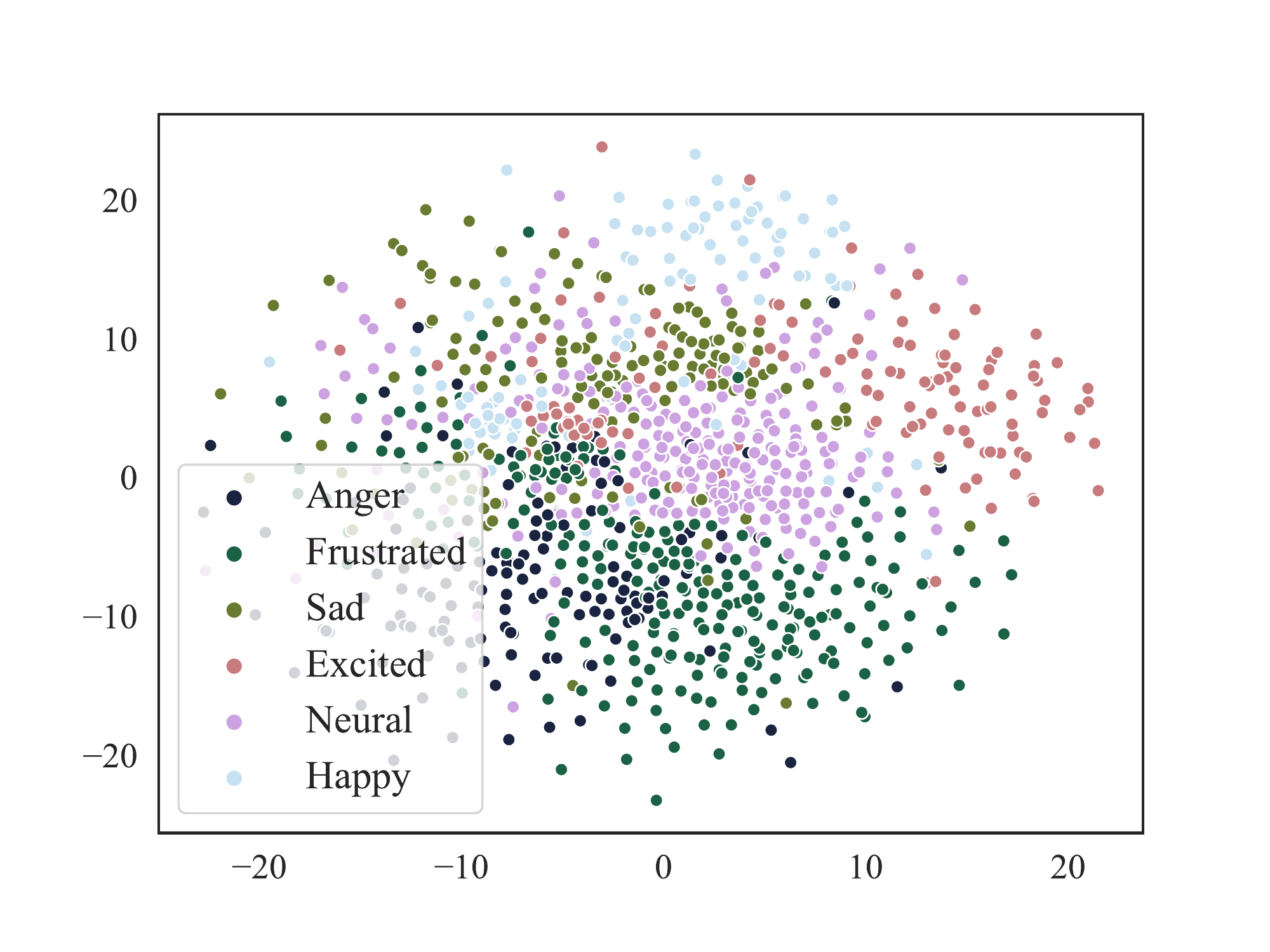}
\end{center}
   \caption{t-SNE result of commom representation of IEMOCAP dataset at missing rate $\eta=0.5$. Clustering of emotions are observed in feature map.}
\label{fig:tsne}
\end{figure}

\noindent\textbf{Common Representation Clustering}\quad The common representation of IEMOCAP dataset at missing rate $\eta=0.5$ is shown in \ref{fig:tsne}, where t-SNE is exploited for visualization. As can be observed, the common representation presents obvious clustering features and margins, especially for sad emotions, which lie at the lower right corner of the scatter map.

\subsection{Ablation Study}
This paper's main novelty is the introduction of party attention and M2R2 framework. We remove these two components once a time and test on the IEMOCAP dataset at missing rate $\eta=0.5$ to study the influence on them, where the results are shown in Table \ref{tbl:ablation}.

As expected, the proposed two components improve the accuracy and F1 score by about 3\% percent in total. In addition, the M2R2 is more influential than party attention, where the accuracy falls about 2.20\% than 0.9\%. 

\begin{table}
\centering
\resizebox{0.9\linewidth}{!}{
\begin{tabular}{|cc|cc|}
\hline
\multicolumn{2}{|c|}{Component} & \multicolumn{2}{c|}{Metrics} \\ \hline
Party Attention & M2R2 Framework & Accuracy & F1 score \\ \hline
\XSolidBrush & \XSolidBrush & 57.82 & 57.43 \\
\CheckmarkBold & \XSolidBrush & 58.68 & 58.16  \\
\CheckmarkBold & \CheckmarkBold & 60.01 & 59.63  \\ \hline
\end{tabular}}
\caption{Ablation results on IEMOCAP at missing rate $\eta=0.5$.}
\label{tbl:ablation}
\end{table}

\section{Conclusions}
We proposed the PANet model and M2R2 Framework for ERC task with missing modalities at the utterance level. The proposed PANet model based on RNN can handle incomplete utterances by spreading dependency on all conversation parties. Our algorithm learns PANet and CPM-Nets of incomplete data jointly, where extended datasets with hidden features for CPM-Nets and common representation for PANet can provide more versatile information for ERC task. We empirically validate that the proposed algorithm is relatively robust to sequential data with missing modalities. More structures and techniques with suitable common representation learning methods should be tested, which we plan to explore in the future.

{\small
\bibliographystyle{ieee_fullname}
\bibliography{main}

\begin{thebibliography}{10}\itemsep=-1pt

\bibitem{reviewmultimodal}
Sharmeen M Saleem~Abdullah Abdullah, Siddeeq Y~Ameen Ameen, Mohammed~AM Sadeeq,
  and Subhi Zeebaree.
\newblock Multimodal emotion recognition using deep learning.
\newblock {\em Journal of Applied Science and Technology Trends}, 2(02):52--58,
  2021.

\bibitem{iemocap}
Carlos Busso, Murtaza Bulut, Chi-Chun Lee, Abe Kazemzadeh, Emily Mower, Samuel
  Kim, Jeannette~N Chang, Sungbok Lee, and Shrikanth~S Narayanan.
\newblock Iemocap: Interactive emotional dyadic motion capture database.
\newblock {\em Language resources and evaluation}, 42(4):335--359, 2008.

\bibitem{reviewercsentimentai}
E. Cambria, Soujanya Poria, Alexander Gelbukh, and Mike~A Thelwall.
\newblock Sentiment analysis is a big suitcase.
\newblock {\em IEEE Intelligent Systems}, 32:74--80, 2017.

\bibitem{gru}
Junyoung Chung, Caglar Gulcehre, KyungHyun Cho, and Yoshua Bengio.
\newblock Empirical evaluation of gated recurrent neural networks on sequence
  modeling.
\newblock {\em arXiv preprint arXiv:1412.3555}, 2014.

\bibitem{vader}
Nasrin Dehbozorgi, Mary~Lou Maher, and Mohsen Dorodchi.
\newblock Sentiment analysis on conversations in collaborative active learning
  as an early predictor of performance.
\newblock In {\em 2020 IEEE Frontiers in Education Conference (FIE)}, pages
  1--9. IEEE, 2020.

\bibitem{reviewreallife}
Laurence Devillers, Laurence Vidrascu, and Lori Lamel.
\newblock Challenges in real-life emotion annotation and machine learning based
  detection.
\newblock {\em Neural Networks}, 18(4):407--422, 2005.
\newblock Emotion and Brain.

\bibitem{SMVAE}
Changde Du, Changying Du, Hao Wang, Jinpeng Li, Wei-Long Zheng, Bao-Liang Lu,
  and Huiguang He.
\newblock Semi-supervised deep generative modelling of incomplete
  multi-modality emotional data.
\newblock In {\em Proceedings of the 26th ACM International Conference on
  Multimedia}, MM '18, page 108–116, New York, NY, USA, 2018. Association for
  Computing Machinery.

\bibitem{opensmile}
Florian Eyben, Martin W{\"o}llmer, and Bj{\"o}rn Schuller.
\newblock Opensmile: the munich versatile and fast open-source audio feature
  extractor.
\newblock In {\em Proceedings of the 18th ACM international conference on
  Multimedia}, pages 1459--1462, 2010.

\bibitem{reviewhci}
Jose~Maria Garcia-Garcia, Victor~MR Penichet, and Maria~D Lozano.
\newblock Emotion detection: a technology review.
\newblock In {\em Proceedings of the XVIII international conference on human
  computer interaction}, pages 1--8, 2017.

\bibitem{cosmic}
Deepanway Ghosal, Navonil Majumder, Alexander Gelbukh, Rada Mihalcea, and
  Soujanya Poria.
\newblock Cosmic: Commonsense knowledge for emotion identification in
  conversations.
\newblock {\em arXiv preprint arXiv:2010.02795}, 2020.

\bibitem{wgan}
Ishaan Gulrajani, Faruk Ahmed, Martin Arjovsky, Vincent Dumoulin, and Aaron
  Courville.
\newblock Improved training of wasserstein gans.
\newblock {\em arXiv preprint arXiv:1704.00028}, 2017.

\bibitem{icon}
Devamanyu Hazarika, Soujanya Poria, Rada Mihalcea, Erik Cambria, and Roger
  Zimmermann.
\newblock Icon: Interactive conversational memory network for multimodal
  emotion detection.
\newblock In {\em Proceedings of the 2018 conference on empirical methods in
  natural language processing}, pages 2594--2604, 2018.

\bibitem{cmn}
Devamanyu Hazarika, Soujanya Poria, Amir Zadeh, Erik Cambria, Louis-Philippe
  Morency, and Roger Zimmermann.
\newblock Conversational memory network for emotion recognition in dyadic
  dialogue videos.
\newblock {\em Proceedings of the conference. Association for Computational
  Linguistics. North American Chapter. Meeting}, 2018:2122---2132, June 2018.

\bibitem{hierarchical}
Chenyang Huang, Amine Trabelsi, and Osmar~R Za{\"\i}ane.
\newblock {\em arXiv preprint arXiv:1904.00132}, 2019.

\bibitem{crosstransformer}
Xincheng Ju, Dong Zhang, Junhui Li, and Guodong Zhou.
\newblock Transformer-based label set generation for multi-modal multi-label
  emotion detection.
\newblock In {\em Proceedings of the 28th ACM International Conference on
  Multimedia}, pages 512--520, 2020.

\bibitem{3dcnn}
Konstantinos Kamnitsas, Christian Ledig, Virginia~FJ Newcombe, Joanna~P
  Simpson, Andrew~D Kane, David~K Menon, Daniel Rueckert, and Ben Glocker.
\newblock Efficient multi-scale 3d cnn with fully connected crf for accurate
  brain lesion segmentation.
\newblock {\em Medical image analysis}, 36:61--78, 2017.

\bibitem{textfeat}
Yoon Kim.
\newblock Convolutional neural networks for sentence classification.
\newblock {\em arXiv: Computation and Language}, 2014.

\bibitem{adam}
Diederik~P. Kingma and Jimmy Ba.
\newblock Adam: A method for stochastic optimization.
\newblock In {\em International Conference on Learning Representations}, 2015.

\bibitem{overviewpartial}
Shao-Yuan Li, Yuan Jiang, and Zhi-Hua Zhou.
\newblock Partial multi-view clustering.
\newblock In {\em Proceedings of the AAAI conference on artificial
  intelligence}, volume~28, 2014.

\bibitem{SMIL}
Mengmeng Ma, Jian Ren, Long Zhao, Sergey Tulyakov, Cathy Wu, and Xi Peng.
\newblock Smil: Multimodal learning with severely missing modality.
\newblock {\em arXiv preprint arXiv:2103.05677}, 2021.

\bibitem{dialoguernn}
Navonil Majumder, Soujanya Poria, Devamanyu Hazarika, Rada Mihalcea, Alexander
  Gelbukh, and Erik Cambria.
\newblock Dialoguernn: An attentive rnn for emotion detection in conversations.
\newblock In {\em Proceedings of the AAAI Conference on Artificial
  Intelligence}, volume~33, pages 6818--6825, 2019.

\bibitem{gantrain}
Lars Mescheder, Andreas Geiger, and Sebastian Nowozin.
\newblock Which training methods for gans do actually converge?
\newblock In {\em International conference on machine learning}, pages
  3481--3490. PMLR, 2018.

\bibitem{igcn}
Weizhi Nie, Rihao Chang, Minjie Ren, Yuting Su, and Anan Liu.
\newblock I-gcn: Incremental graph convolution network for conversation emotion
  detection.
\newblock {\em IEEE Transactions on Multimedia}, 2021.

\bibitem{missing2}
Shigeyuki Oba, Masa aki Sato, Ichiro Takemasa, Morito Monden, Kenichi
  Matsubara, and Shin Ishii.
\newblock A bayesian missing value estimation method for gene expression
  profile data.
\newblock {\em Bioinformatics}, 19:2088--2096, 2003.

\bibitem{MMER}
Mathieu Pag\'{e}~Fortin and Brahim Chaib-draa.
\newblock Multimodal multitask emotion recognition using images, texts and
  tags.
\newblock In {\em Proceedings of the ACM Workshop on Crossmodal Learning and
  Application}, WCRML '19, page 3–10, New York, NY, USA, 2019. Association
  for Computing Machinery.

\bibitem{TrainingST}
Srinivas Parthasarathy and Shiva Sundaram.
\newblock Training strategies to handle missing modalities for audio-visual
  expression recognition.
\newblock {\em arXiv: Audio and Speech Processing}, 2020.

\bibitem{autoencoder}
Duc-Anh Phan, Yuji Matsumoto, and Hiroyuki Shindo.
\newblock Autoencoder for semisupervised multiple emotion detection of
  conversation transcripts.
\newblock {\em IEEE Transactions on Affective Computing}, 12(3):682--691, 2018.

\bibitem{defaffective}
Rosalind~W Picard.
\newblock {\em Affective computing}.
\newblock MIT press, 2000.

\bibitem{bclstm}
Soujanya Poria, Erik Cambria, Devamanyu Hazarika, Navonil Majumder, Amir Zadeh,
  and Louis-Philippe Morency.
\newblock Context-dependent sentiment analysis in user-generated videos.
\newblock In {\em Proceedings of the 55th annual meeting of the association for
  computational linguistics (volume 1: Long papers)}, pages 873--883, 2017.

\bibitem{meld}
Soujanya Poria, Devamanyu Hazarika, Navonil Majumder, Gautam Naik, Erik
  Cambria, and Rada Mihalcea.
\newblock Meld: A multimodal multi-party dataset for emotion recognition in
  conversations.
\newblock {\em arXiv preprint arXiv:1810.02508}, 2018.

\bibitem{reviewerc}
Soujanya Poria, Navonil Majumder, Rada Mihalcea, and Eduard Hovy.
\newblock Emotion recognition in conversation: Research challenges, datasets,
  and recent advances.
\newblock {\em IEEE Access}, 7:100943--100953, 2019.

\bibitem{interactive}
Minjie Ren, Xiangdong Huang, Xiaoqi Shi, and Weizhi Nie.
\newblock Interactive multimodal attention network for emotion recognition in
  conversation.
\newblock {\em IEEE Signal Processing Letters}, 28:1046--1050, 2021.

\bibitem{birnn}
Mike Schuster and Kuldip~K Paliwal.
\newblock Bidirectional recurrent neural networks.
\newblock {\em IEEE transactions on Signal Processing}, 45(11):2673--2681,
  1997.

\bibitem{reviewml}
Ch. Sekhar, M.~Srinivasa Rao, A.~S.~Keerthi Nayani, and Debnath Bhattacharyya.
\newblock Emotion recognition through human conversation using machine learning
  techniques.
\newblock In Debnath Bhattacharyya and N. Thirupathi~Rao, editors, {\em Machine
  Intelligence and Soft Computing}, pages 113--122, Singapore, 2021. Springer
  Singapore.

\bibitem{bayes}
Muhammad Yusril~Helmi Setyawan, Rolly~Maulana Awangga, and Safif~Rafi Efendi.
\newblock Comparison of multinomial naive bayes algorithm and logistic
  regression for intent classification in chatbot.
\newblock In {\em 2018 International Conference on Applied Engineering (ICAE)},
  pages 1--5. IEEE, 2018.

\bibitem{fuzzy}
Catherine Soladi{\'e}, Hanan Salam, Catherine Pelachaud, Nicolas Stoiber, and
  Renaud S{\'e}guier.
\newblock A multimodal fuzzy inference system using a continuous facial
  expression representation for emotion detection.
\newblock In {\em Proceedings of the 14th ACM international conference on
  Multimodal interaction}, pages 493--500, 2012.

\bibitem{missing1}
Olga~G. Troyanskaya, Michael~N. Cantor, Gavin Sherlock, Patrick~O. Brown,
  Trevor Hastie, Robert Tibshirani, David Botstein, and Russ~B. Altman.
\newblock Missing value estimation methods for dna microarrays.
\newblock {\em Bioinformatics}, 17:520--525, 2001.

\bibitem{attention}
Ashish Vaswani, Noam Shazeer, Niki Parmar, Jakob Uszkoreit, Llion Jones,
  Aidan~N Gomez, {\L}ukasz Kaiser, and Illia Polosukhin.
\newblock Attention is all you need.
\newblock In {\em Advances in neural information processing systems}, pages
  5998--6008, 2017.

\bibitem{overviewmissing}
Johannes Wagner, Florian Lingenfelser, and Elisabeth Andr{\'e}.
\newblock Building a robust system for multimodal emotion recognition.
\newblock {\em Emotion recognition: A pattern analysis approach}, pages
  379--410, 2015.

\bibitem{missing3}
Lei Yuan, Yalin Wang, Paul~M. Thompson, Vaibhav~A. Narayan, and Jieping Ye.
\newblock Multi-source learning for joint analysis of incomplete multi-modality
  neuroimaging data.
\newblock In {\em Knowledge Discovery and Data Mining}, 2012.

\bibitem{partial}
Changqing Zhang, Yajie Cui, Zongbo Han, Joey~Tianyi Zhou, Huazhu Fu, and
  Qinghua Hu.
\newblock Deep partial multi-view learning.
\newblock {\em IEEE transactions on pattern analysis and machine intelligence},
  2020.

\bibitem{gcn}
Dong Zhang, Liangqing Wu, Changlong Sun, Shoushan Li, Qiaoming Zhu, and Guodong
  Zhou.
\newblock Modeling both context-and speaker-sensitive dependence for emotion
  detection in multi-speaker conversations.
\newblock In {\em IJCAI}, pages 5415--5421, 2019.

\bibitem{MMIN}
Jinming Zhao, Ruichen Li, and Qin Jin.
\newblock Missing modality imagination network for emotion recognition with
  uncertain missing modalities.
\newblock In {\em Meeting of the Association for Computational Linguistics},
  2021.

\bibitem{knowledge}
Peixiang Zhong, Di Wang, and Chunyan Miao.
\newblock Knowledge-enriched transformer for emotion detection in textual
  conversations.
\newblock {\em arXiv preprint arXiv:1909.10681}, 2019.

\end{thebibliography}
}

\end{document}